\documentclass[12pt, a4paper]{article}

\usepackage [english]{babel}
\usepackage[dvips]{graphicx}

\date{}

\title{Some Aspects of Wave and Quantum Approaches at Description of Movement of Twisted Light}
\author{Yu.A. Portnov\footnote{portnovyura@yandex.ru} \\ {\it Physics Department,} \\
{\it The Moscow state automobile and road technical university
(MADI),} \\ {\it Leningradskiy Prosp. 64, Moscow, 125319,
Russian}}

\begin{document}

\maketitle

\begin{abstract}
The existence of twisted light may be inferred from modern quantum
concepts and experimental data. These waves possess energy,
impulse and angular momentum. However, the Maxwell's
four-dimensional theory of electromagnetism does not imply the
existence of waves with these properties. This article develops a
model generalizing the theory of electromagnetism in such a way
that it would be possible to obtain equations of twisted
electromagnetic waves. Generalization is implemented by
introduction of a space-time with a more complex structure
compared to the four-dimensional space-time. Such spaces include a
seven-dimensional space-time, which allows to describe not only
translational, but also rotational motion of bodies. A model
developed by the author provides the following results: 1)
generalization of the theory of electromagnetism in which it is
possible to obtain equations of twisted light waves, 2) solution
describing interference of light waves oppositely twisted, 3) the
formula relating the energy, impulse and angular momentum of
electromagnetic wave, 4) justification of a new phenomenon -
redshift due to electromagnetic waves screwing.
\end{abstract}

{\bf UDK:} 531-4:537.877



\date{12/I/2015} 

\section{Introduction}

Any light wave can carry not only energy $E=\hbar\omega$ and an
impulse $p=k\hbar$, but an angular momentum, as well. The angular
momentum of light consists of two components: a spin angular
momentum and an orbital angular momentum. The spin angular
momentum $L_{s}=l\hbar$ is related to wave polarization, and the
orbital angular momentum $L_{o}=n\hbar$ is related to dependence
of a wave phase from direction, where $l$, $n$ are some whole
quantum numbers.

\begin{figure}[h]
\begin{minipage}[h]{0.19\linewidth}
\center{\includegraphics[width=1\linewidth]{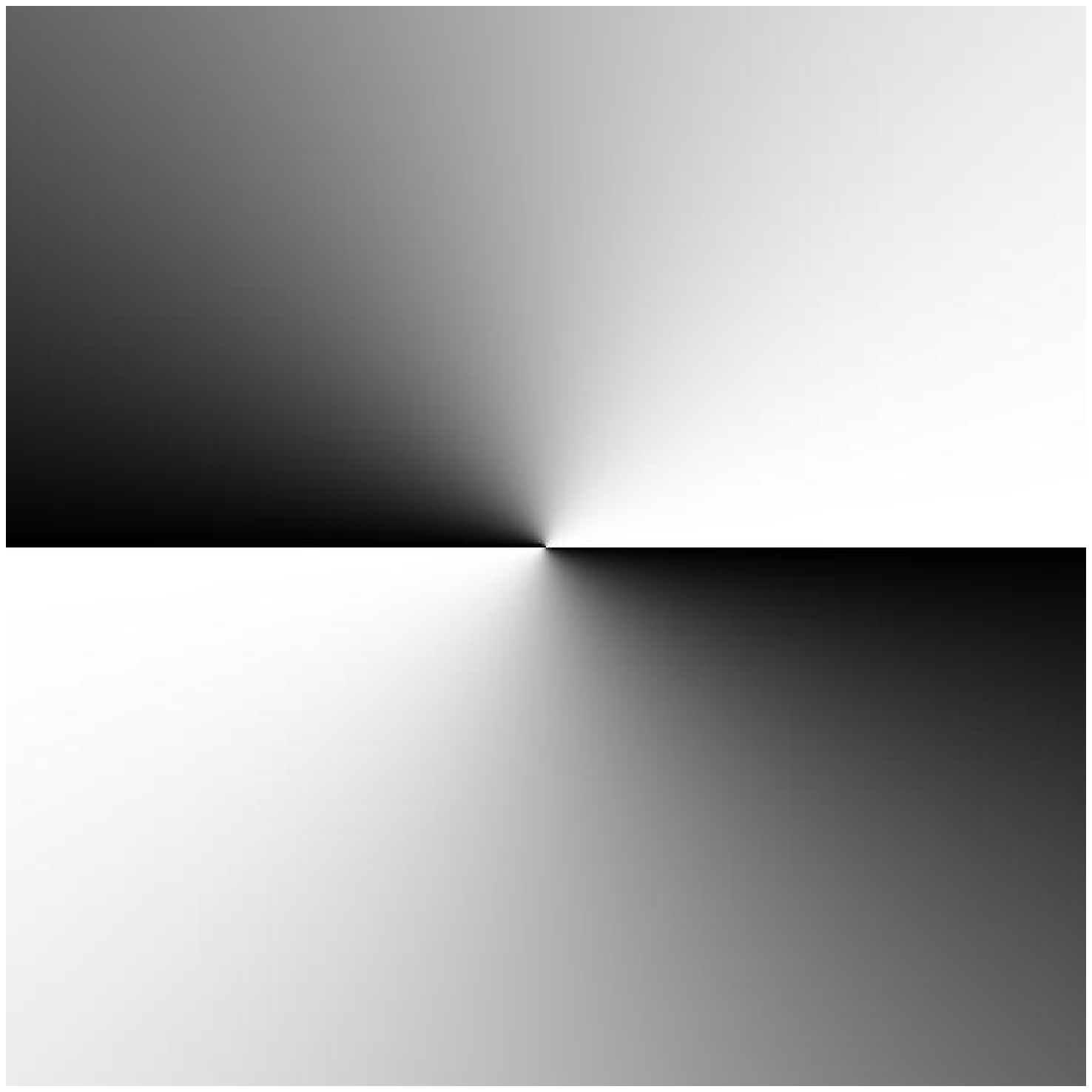}} \\ a)
\end{minipage}
\begin{minipage}[h]{0.19\linewidth}
\center{\includegraphics[width=1\linewidth]{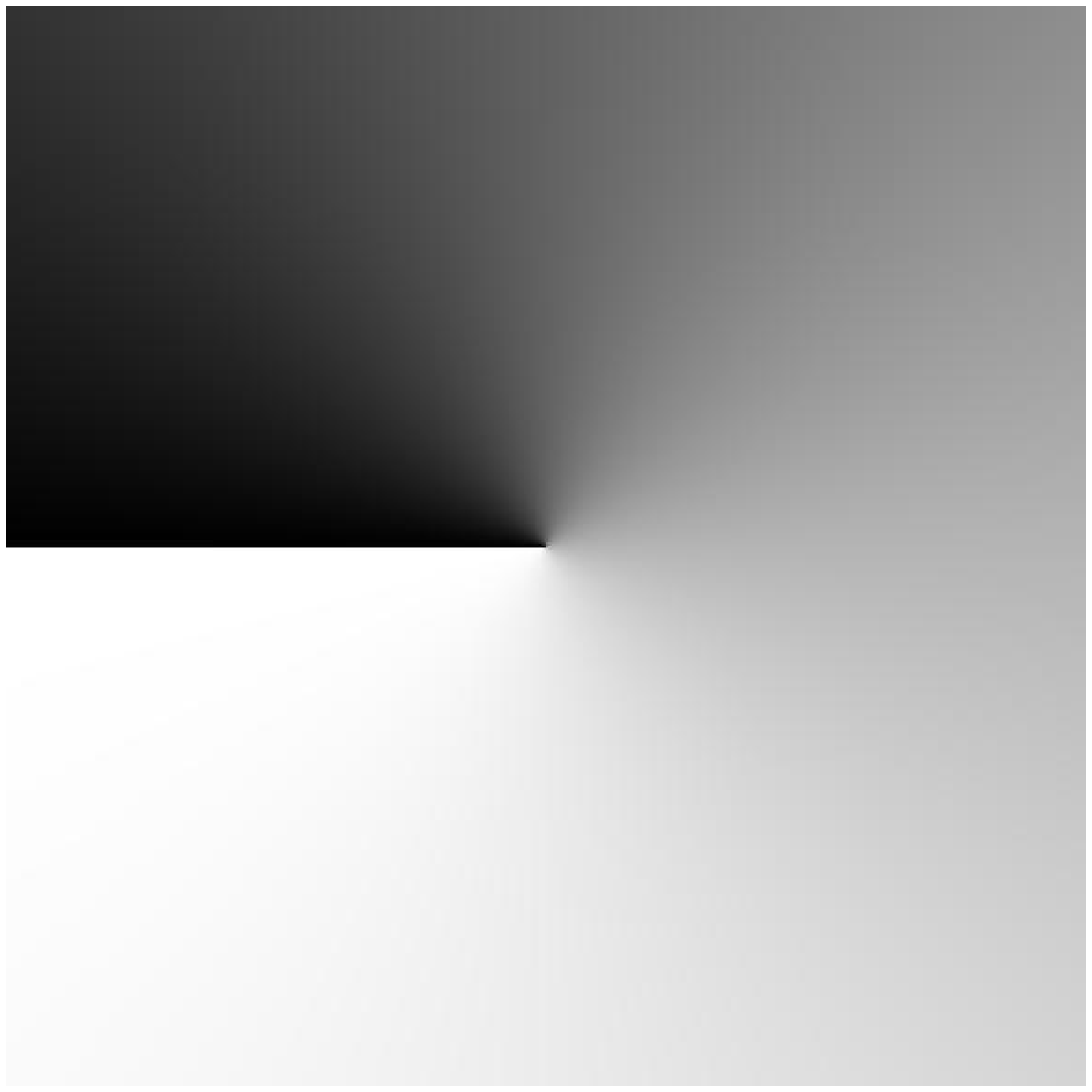}}
\\ b)
\end{minipage}
\hfill
\begin{minipage}[h]{0.19\linewidth}
\center{\includegraphics[width=1\linewidth]{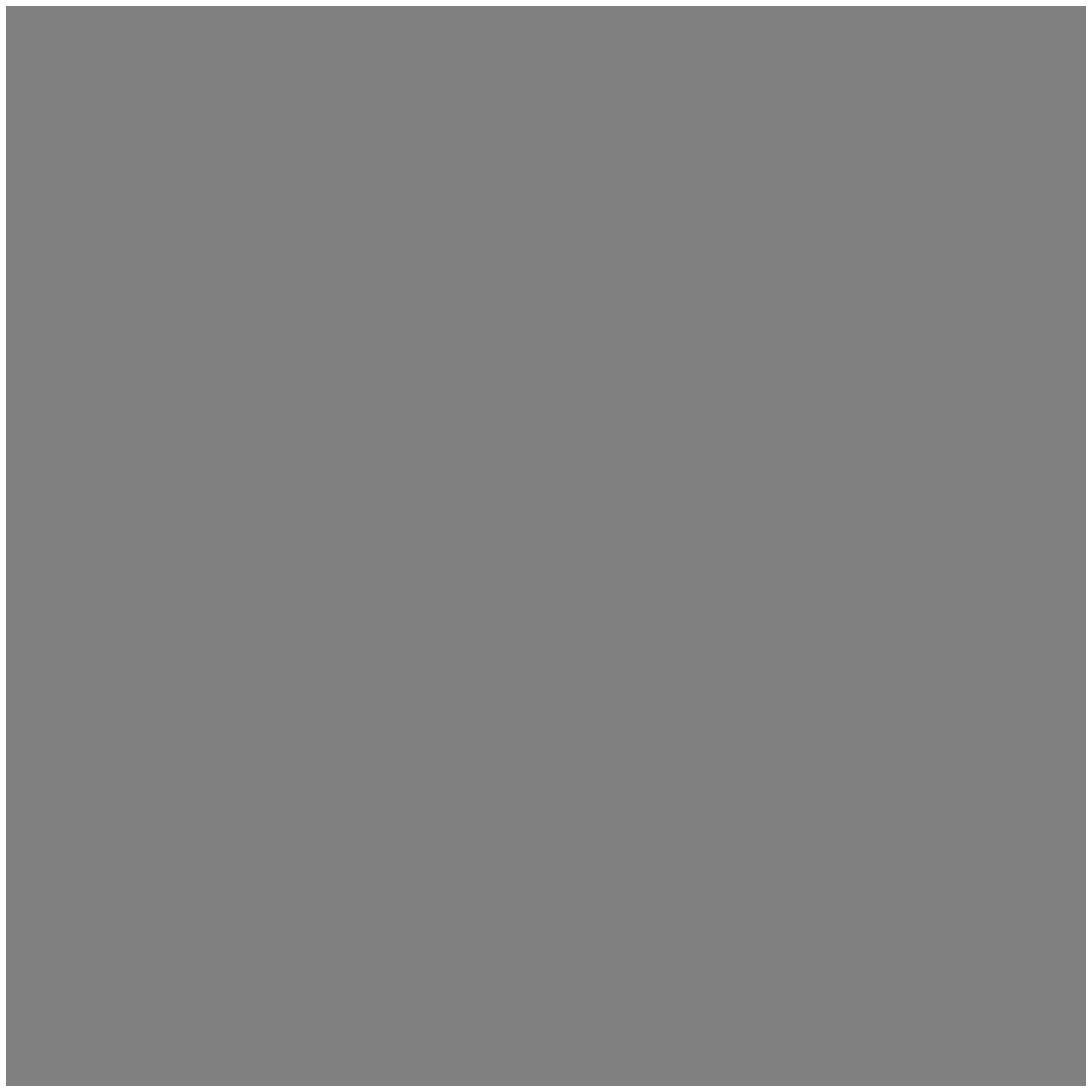}}
\\ c)
\end{minipage}
\hfill
\begin{minipage}[h]{0.19\linewidth}
\center{\includegraphics[width=1\linewidth]{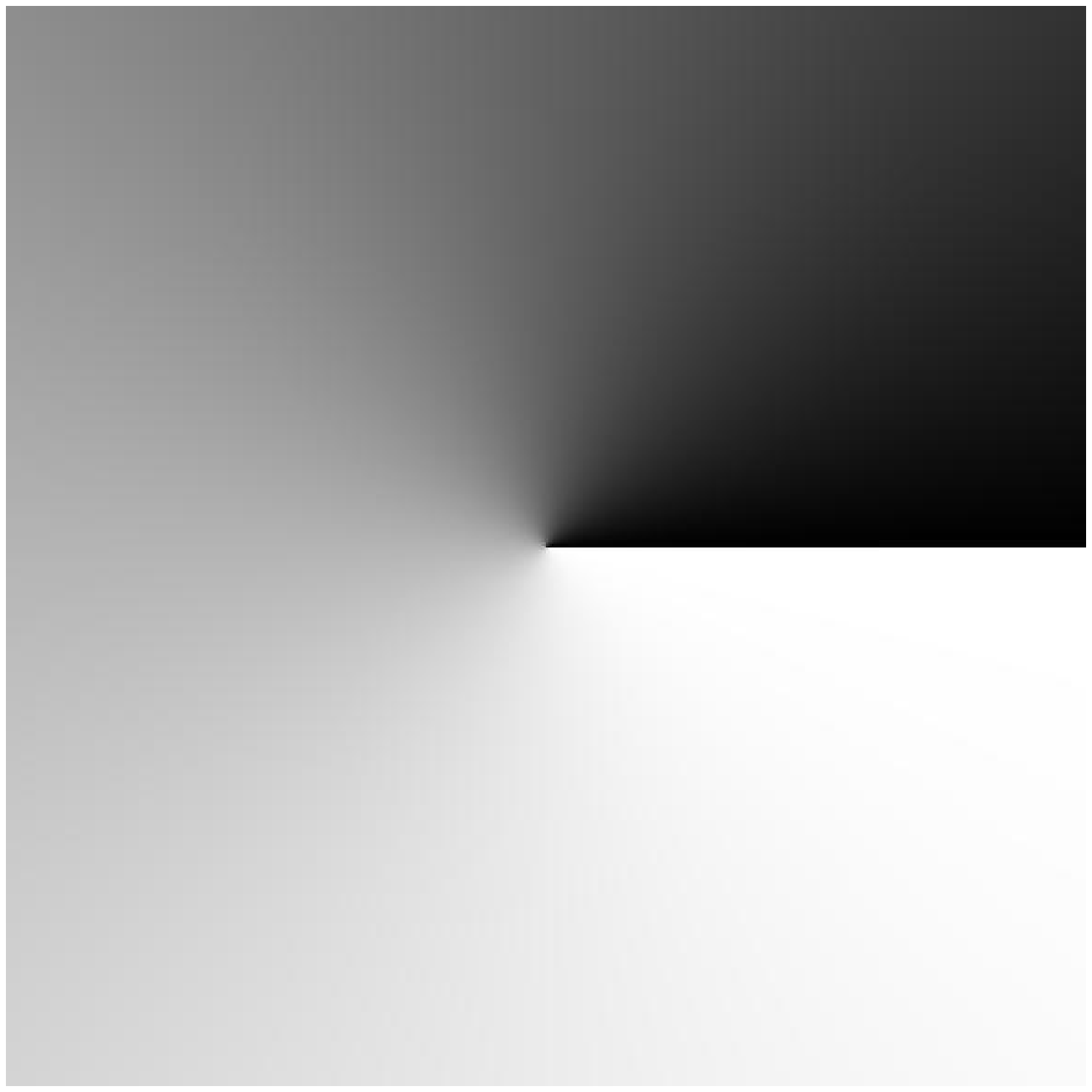}}
\\ d)
\end{minipage}
\begin{minipage}[h]{0.19\linewidth}
\center{\includegraphics[width=1\linewidth]{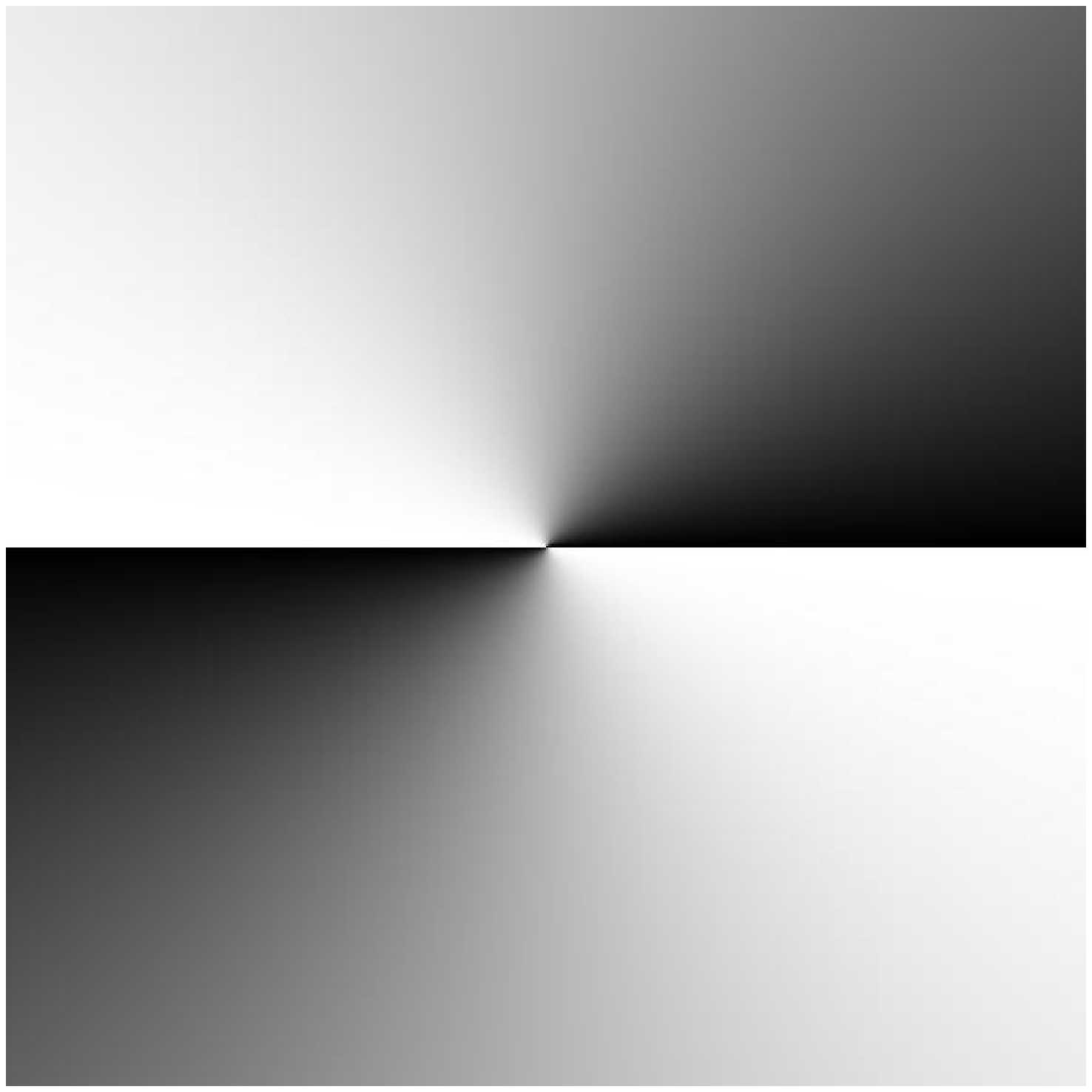}}
\\ e)
\end{minipage}
\caption{A phase profile of a light ray, with quantum numbers: a)
$n=-2$, b) $n=-1$, c) $n=0$, d) $n=+1$, e) $n=+2$. The wave phase
equal to zero is colored black, the wave phase equal to $2\pi$ is
colored white.} \label{fig1}
\end{figure}

An electromagnetic wave is characterized by two directions: a wave
vector $\vec{k}$ and an electric-field vector $\vec{E}$ constantly
being mutually perpendicular. As a result, another degree of
freedom occurs, i. e. rotation of the electric-field vector around
the wave vector. With such rotation change of the electric-field
vector shows as a phenomena of polarization. In a common plane
light wave $n=0$ all wavefronts follow each other. Therefore, if
we take a wave cross-section, the wave phase will be equal in each
point of the cross-section (Fig. \ref{fig1}с).

The work by Allen L. et al. \cite{portnov:bib:Allen}, published in
1992, proposed schemes for formation and detection of twisted
light, and was the first to suggest that the electromagnetic wave
had the orbital angular momentum apart from the spin moment. In
the twisted light the wavefront resembles a spiral directed
towards the side of wave propagation with the phase changing in
each cross-sectional point depending on the direction. For
instance, at $n=\pm1$ (Fig. \ref{fig1}(b,d)) the wave phase
changes by $2\pi$ for one turn of the wavefront, and at $n=\pm2$
(Fig. \ref{fig1}(a,e)) for a half-turn.

\begin{figure}[h]
\begin{minipage}[h]{0.48\linewidth}
\center{\includegraphics[width=1\linewidth]{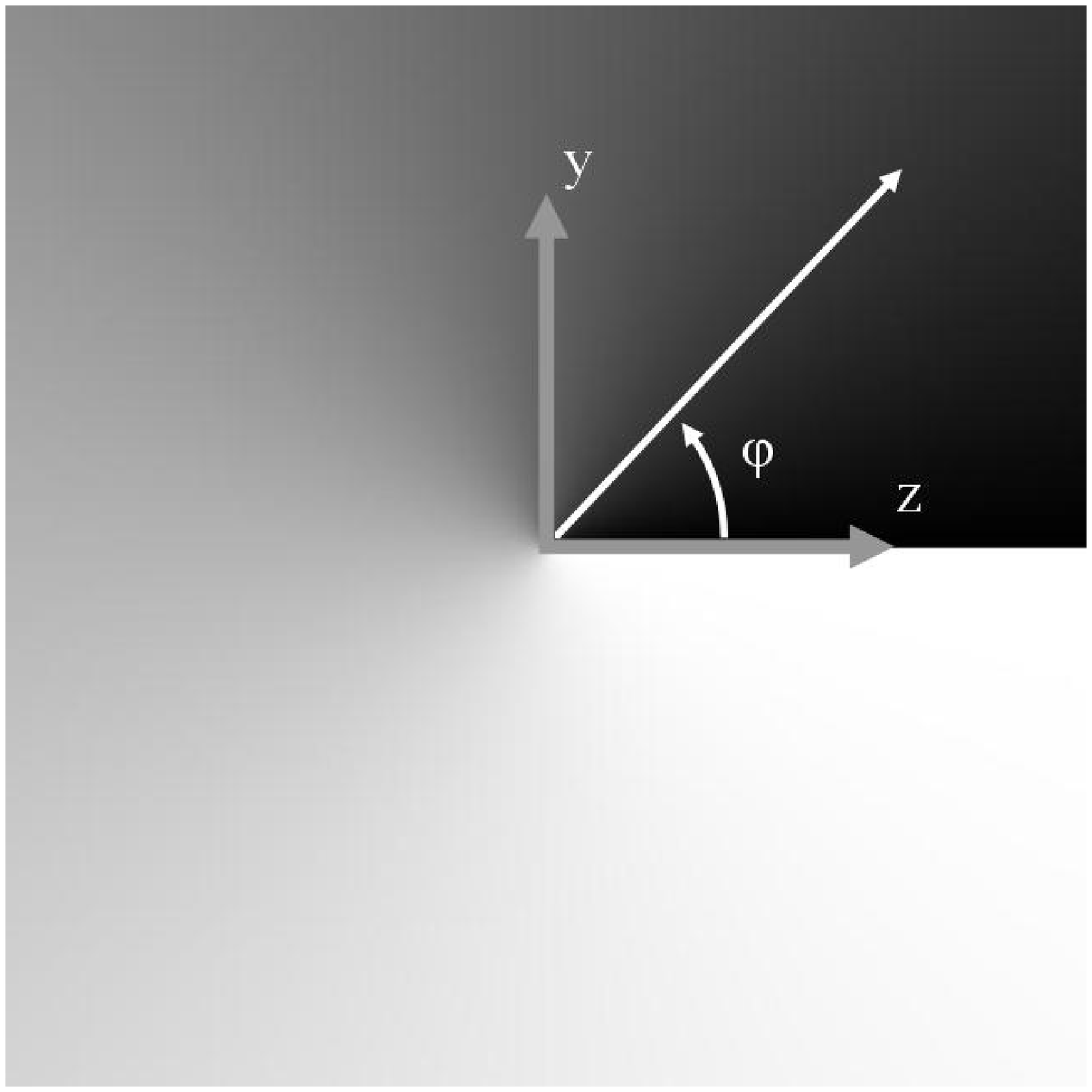}}
\\ a)
\end{minipage}
\begin{minipage}[h]{0.48\linewidth}
\center{\includegraphics[width=1\linewidth]{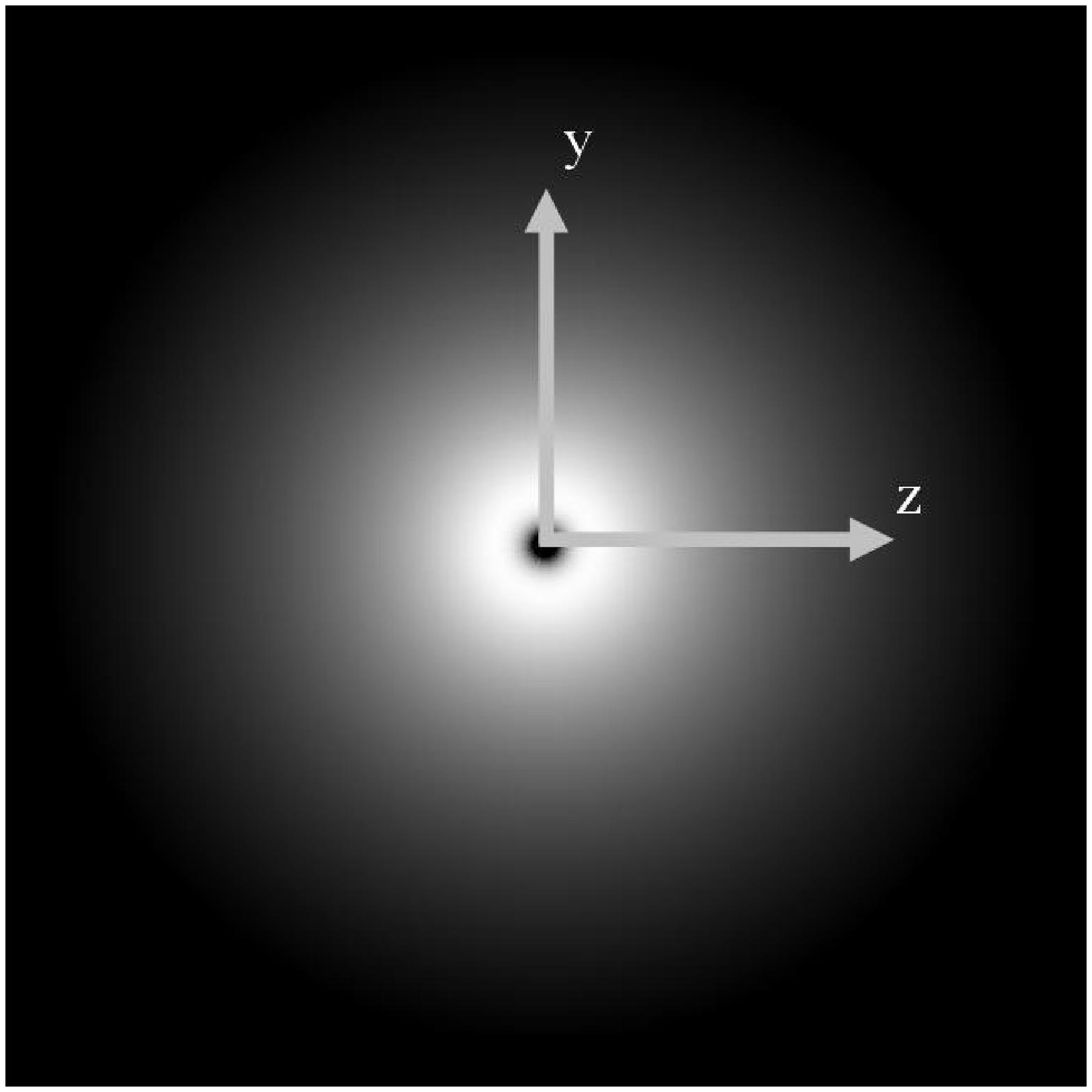}}
\\ b)
\end{minipage}
\caption{a) Phase profile of the light beam with $n=+1$ depending
on the turning angle of the wavefront $\varphi$. In the figure,
the wave phase equal to zero is colored black, the wave phase
equal to $2\pi$ is colored white. b) The amplitude of the twisted
light beam; maximum amplitude is colored white, while the minimum
one is colored black.} \label{fig2}
\end{figure}

Straight twisted light beams were experimentally generated by
\cite{portnov:bib:He} in 1995. Twisted electromagnetic waves were
formed technically with a specific prism of variable thickness.
Difference in thickness of the prism makes it possible for the
wave phase, which has passed through the thicker prism layer, to
lag from the wave phase, which has passed through the less thick
prism layer. Thus, at the output different wave sections have a
different phase depending on the direction (Fig. \ref{fig2}a),
while the amplitude attains the distribution Fig. \ref{fig2}b. The
advanced methods for generating twisted light allow to
simultaneously form several beams differently twisted. The work
\cite{portnov:bib:Benjamin} used a pitch-fork hologram for these
purposes.

An experiment described in the work \cite{portnov:bib:Arita}
proves that the light wave possess not only the impulse, but the
orbital angular momentum, as well. The authors describe an
experiment, where a microparticle was suspended in the focus of a
laser beam. When having absorbed light, the microparticle started
turning. Thereby, the turning direction depended on twisting
direction of the laser light.

According to the work \cite{portnov:bib:Bahrdt}, it was possible
to generate twisted x-ray waves with photon energy equal to 99 eV
due to developed technologies for formation and registration of
the twisted light.

The twisted light has various applications today, including the
quantum information theory, microcomputer control and
astrophysical researches \cite{portnov:bib:Torres}. For instance,
in 2010 B. Thide with the colleagues issued the work
\cite{portnov:bib:Tamburini}, describing a method allowing
determination of rotation features of black holes based on the
analysis of an angular momentum of the light passing near an
accretion disc.

The simplified equation of the twisted light wave propagating
along the axis $Ox$, with the phase depending on time $t$,
component $x$ and the wavefront rotation angle $\varphi$ (Fig.
\ref{fig2}a), is of the following form:
\begin{eqnarray*}
E=E_{0}\cdot\exp\left(-i\left(\omega t-kx-n\varphi\right)\right).
\end{eqnarray*}
However this equation does not flows from the Maxwell's
four-dimensional theory of electromagnetism, suggesting that the
standard theory of electromagnetic fields shall be corrected.

\section{Choice of Description Method for Twisted Waves}

Therefore, the purpose of this work is to build a model of
electromagnetism, naturally involving a possibility to describe
twisted electromagnetic waves, and to investigate the properties
of the built model.

Since the electromagnetic wave in a three-dimensional space has
additional degrees of freedom, we may suppose that to describe
movement of the twisted electromagnetic waves it is reasonable to
use a space-time with a more complex structure, rather than the
four-dimensional space-time.

The works by mathematicians B. Riemann, G. Weyl, E. Cartan, J.
Schouten and others show that spaces may be characterized by
curvature, as well as by torsion and nonmetricity. For instance,
to describe the observed phenomena the modern cosmology uses the
Riemann-Cartan space with curvature and torsion or a more common
affine-metric space with curvature, torsion and nonmetricity,
particularly the Weyl-Cartan space with the nonmetricity of Weyl
type \cite{portnov:bib:BabLipFrol}-\cite{portnov:bib:Portnov2015}.
However all attempts of G. Weyl to use the concepts of scale
conversion with parallel translation to create a geometrical
theory of electromagnetism failed that makes us to decline the
concepts of development of a generalized theory of
electromagnetism with the use of spaces with torsion and
nonmetricity.

Another example of the space-time accounting for additional
degrees of freedom is a model of the seven-dimensional space-time.
As the works
\cite{portnov:bib:Portnov}-\cite{portnov:bib:PortnovRAP} show, to
explain the dynamics not only of translational motion, but of
rotational motion, as well, of bodies in gravitational fields, one
may use the seven-dimensional space-time which includes, apart
from time and three spatial components, three components aligning
the body in the space $x^{4}=\varphi$, $x^{5}=\psi$,
$x^{6}=\theta$, the so called Euler angles. The equations of the
geodesic seven-dimensional space-time allow to develop not only
equations for translational motion, but equations describing
rotational motion of gyroscopes \cite{portnov:bib:PortnovRAP}, as
well.

The metric tensor takes the following form for the inane
seven-dimensional space-time
\cite{portnov:bib:Portnov}-\cite{portnov:bib:PortnovRAP}:
$$g_{00}=-g_{\alpha\alpha}=1,$$
\begin{equation}
\label{e1}   g_{45}=g_{54}=-\frac{J_{\omega}\cos(\theta)}{m},
\end{equation}
$$g_{44}=g_{55}=g_{66}=-\frac{J_{\omega}}{m},$$ where $J_{\omega}$
- the angular momentum of a test body relating to the rotation
axis, $m$ - the weight of the test body, $\alpha=1,2,3$. The
paradox of notation of the metric tensor
\cite{portnov:bib:PortnovGC2014}-\cite{portnov:bib:PortnovST2014},
which, regardless of four-dimensional metric tensors of the
general relativity, depends on the parameters of the test body
(inertia momentum to weight ratio), makes it possible to suppose
that the space is not an absolute value and the gravitational
field depends on the test body inserted into it. However it is
important that deflection of geodesic lines, along which the test
body move, is the only method available today to detect
gravitation. That is why we may suggest about presence or lack of
gravitation only in the view of the moving test body. As a result,
it is no wonder that the metric tensor describing the
gravitational field will depend on the parameters of the test
body. This concept makes to reconsider the notion of relativity:
not only motion becomes relative, but the space-time itself
depends on the viewed test body.

As was shown in the works
\cite{portnov:bib:Portnov}-\cite{portnov:bib:PortnovST2014}, the
use of seven-dimensional space-time for description of rotational
motion of bodies proved to be successful. This includes motion of
gyroscopes in gravitational fields, description of motion of the
galactic disc related to stellar rotation, as well as change of
rotational speed of the bodies with the change of the
gravitational potential. From the above it can be concluded that
the model of the seven-dimensional space-time shall be used to
develop the generalized theory of electromagnetism.

\section{Derivation of Twisted Waves Equations}

We shall generalize the Maxwell's four-dimensional theory of
electromagnetism by the space-time of seven dimensions to describe
the twisted electromagnetic waves. Interaction of particles is
described by a field of force with properties, regardless of the
classical theory, characterized by a 7-vector $A_{i}$. Further it
will be called a7-potential with the components being functions of
the coordinates, time and angles of orientation
\cite{portnov:bib:Portnov}. Three spatial components of the
7-potential $A^{k}$ form a three-dimensional vector, called a
field rotational potential, a time component will be called a
scalar potential $A^{0}=\Phi$, and three orientating components of
the 7-potential will form the field rotational potential. The
index of the 7-potential $A^{k}$ will be lowered by the metric
tensor (\ref{e1}) $$A_{i}=g_{ik}A^{k}.$$

The Lagrange function for a charged body in the electromagnetic
field takes the following form for the seven-dimensional
space-time:
\begin{equation}
\label{e4}
L=-mc^{2}\sqrt{1-\frac{V^{2}}{c^{2}}-\frac{J_{\omega}\omega^{2}}{mc^{2}}}-q\Phi+\frac{q}{c}(A_{i}V^{i})+\frac{q}{c}(A_{j}\omega^{j}).
\end{equation}
Therefore, it is important that due to uniformity and isotropy of
the space it may be supposed that the rotational part of the
potential of the electromagnetic field depends only on the angle
coordinates, while the dimensional part depends on the linear
coordinates.

The Lagrange equations determine the equations of the charge
motion in the given electromagnetic space:
\begin{equation}
\label{e7} \frac{d}{dt}\frac{\partial L}{\partial
u^{i}}-\frac{\partial L}{\partial x^{i}}=0,
\end{equation}
where $L$ is defined by the formula (\ref{e4}). The velocity
derivative is a generalized impulse of the body $p_{i}=\partial
L/\partial u^{i}$. Then the motion equation will be as follows:
\begin{equation}
\label{e8}
\frac{d}{dt}\left(p_{k}+\frac{q}{c}A_{k}\right)=\frac{q}{c}\partial_{k}\left(-c\Phi+(A_{n}u^{n})\right),
\end{equation}
where $k,n=1,2,3,4,5,6$. Following notation of the total
differential of the 7-potential and transformation, the following
will be derived:
\begin{equation}
\label{e9} \frac{d p_{k}}{dt}=-\frac{q}{c}\frac{\partial
A_{k}}{\partial t}-q
\partial_{k}\Phi+\frac{q}{c}\left(u^{n}\varepsilon_{knl}\varepsilon_{sdh}g^{ls}g^{dm}g^{kf}\partial_{m}A_{f}\right),
\end{equation}
where $\varepsilon_{knl}$ - symbols similar to the Levi-Civita
symbols \cite{portnov:bib:Hermann}, with the coefficients running
the values from 1 through 6. Since dimensional and angular
coordinates may not be multiplied in the considered space, the
Levi-Civita symbols will have the following structure:
\begin{equation}
\label{e10} \varepsilon_{ikl}=\left\{
 \begin{array}{cc}
 -1 & (1,2,3); (2,3,1); (3,1,2); (4,5,6); (5,6,4); (6,4,5) \\
 +1 & (3,2,1); (1,3,2); (2,1,3); (6,5,4); (4,6,5); (5,4,6) \\
  0 &
 \end{array} \right\}
\end{equation}

In the left-hand parts of the equations (\ref{e9}) there is a time
derivative of the body impulse. Therefore, in the right-hand parts
of the equations there is a force impacting the body. By analogy
with the theory of electromagnetism, the force impacting the body
will be divided into two parts: one depending on the velocity and
another not depending.

The force of the first type is the electric field:
\begin{equation}
\label{e11} E_{k}=-\frac{1}{c}\frac{\partial A_{k}}{\partial
t}-\partial_{k}\Phi.
\end{equation}
The force of the second type is the magnetic field:
\begin{equation}
\label{e12} H_{r}=\varepsilon_{rdh}g^{dm}g^{hf}\partial_{m}A_{f}.
\end{equation}
By combining the parametric equations (\ref{e11}) and (\ref{e12})
the first Maxwell equation modified for the seven-dimensional case
may be developed:
\begin{equation}
\label{e14}
\varepsilon_{rdh}g^{dm}g^{hf}\partial_{m}E_{f}=-\frac{1}{c}\frac{\partial}{\partial
t}H_{r}.
\end{equation}
By multiplying (\ref{e12}) by a seven-dimensional nabla operator
$\nabla=\partial_{k}dx^{k}$ the second modified Maxwell equation
may be established:
\begin{equation}
\label{e15} g^{kh}\partial_{k}H_{h}=0.
\end{equation}

The product of the charge density $\varepsilon$ by the
seven-dimensional velocity vector $u^{k}$
\cite{portnov:bib:Portnov} will be called a current density
7-vector:
\begin{equation}
\label{e20} j^{k}=\varepsilon u^{k}.
\end{equation}
Three spatial components thereof form three-dimensional current
density, while the rotational components form charge rotational
density.

A seven-dimensional tensor of the electromagnetic field will be
introduced:
\begin{equation}
\label{e17} F_{hk}=\partial_{h}A_{k}-\partial_{k}A_{h}.
\end{equation}
By locating the field equation by the principle of least action
the following equation may be developed:
\begin{equation}
\label{e25} \partial_{k}F^{ik}=-\frac{4\pi}{c}j^{i},
\end{equation}
which is the second pair of the Maxwell equations noted in the
seven-dimensional form. By substituting different values $i$ and
tensor components (\ref{e17}) the following equations will be
established:
\begin{equation}
\label{e26} g^{km}\partial_{k}E_{m}=4\pi\varepsilon,
\end{equation}
\begin{equation}
\label{e27}
\varepsilon_{klm}g^{lh}g^{mf}\partial_{h}H_{f}=\frac{1}{c}\frac{\partial
E_{k}}{\partial t}+\frac{4\pi}{c}j_{k},
\end{equation}
The equations (\ref{e15}), (\ref{e14}) alongside with the
equations (\ref{e26}), (\ref{e27}) define the electromagnetic
field in the seven-dimensional space. Further these equations will
be called the Maxwell seven-dimensional equations.

To establish equations of electromagnetic wave the Maxwell
seven-dimensional equations in the inane space-time will be
considered, i. e., $\varepsilon=0$ и $j_{k}=0$. Let us take a
seven-dimensional rotation of the equation (\ref{e14}):
$$\varepsilon_{eru}g^{rb}g^{us}\partial_{b}\varepsilon_{sdh}g^{dm}g^{hf}\partial_{m}E_{f}=-\frac{1}{c}\frac{\partial}{\partial
t}\varepsilon_{eru}g^{rb}g^{us}\partial_{b}H_{s}.$$ By using the
equation (\ref{e27}) a wave equation of the following type may be
noted:
\begin{equation}
\label{e36}
\varepsilon_{eru}\varepsilon_{sdh}g^{rb}g^{us}g^{dm}g^{hf}\partial_{b}\partial_{m}E_{f}=-\frac{1}{c^{2}}\frac{\partial^{2}E_{e}}{\partial
t^{2}}.
\end{equation}
The product $\varepsilon_{eru}\varepsilon^{mfu}$ is the true
sixth-rank tensor \cite{portnov:bib:Dimitrienko}, which may
expressed as a combination of product of components of an identity
tensor. A proposal that the functions of time, dimensional and
rotational coordinates are components of the electrical induction
vector $E=E^{2}(t,x,\varphi)$ allows to significantly simplify the
equation (\ref{e36}) as follows:
\begin{equation}
\label{e38}
\left(\partial_{1}^{2}+\frac{1}{R_{in}^{2}}\partial_{4}^{2}\right)E+\frac{1}{c^{2}}\frac{\partial^{2}}{\partial
t^{2}}E=0.
\end{equation}
where $R_{in}$ - a radius of inertia. The equation (\ref{e38}) in
case of motion of motion of the wave along the axis $Ox$ and
rotation around this axis with the phase turn by angle $\varphi$,
will be solved as follows:
\begin{equation}
\label{e39} E=A\cdot\exp\left(-i\left(\omega
t-\frac{2\pi}{\lambda}x-\frac{2\pi}{\mu}\varphi+\sigma\right)\right),
\end{equation}
where $\omega$ - a cyclical oscillation frequency; $\lambda=cT$ -
the wave length; $\mu=\Omega T$ - will be called a wave rotation
angle, it is the angle by which the wave phase rotates per one
time period; $\Omega$ - an angular velocity of the wave phase
rotation; $\sigma$ - an initial oscillation phase. Therefore the
developed equation (\ref{e39}) is the equation of the twisted
electromagnetic wave with the phase changing depending on the
time, coordinate and rotation angle.

\section{Twisted Waves Interference}

Let us take two twisted electromagnetic waves $E'$ and $E''$ with
equal amplitudes, equal wavelengths and equal in modulus rotation
angles $\mu_{1}=-\mu_{2}=\mu$ are distributed along the axis $Ox$:
\begin{equation}
\label{e42} E'=A\cdot\exp\left(-i\left(\omega
t-\frac{2\pi}{\lambda}x-\frac{2\pi}{\mu}\varphi+\sigma_{1}\right)\right),
\end{equation}
\begin{equation}
\label{e43} E''=A\cdot\exp\left(-i\left(\omega
t-\frac{2\pi}{\lambda}x+\frac{2\pi}{\mu}\varphi+\sigma_{2}\right)\right).
\end{equation}
Interference of the waves (\ref{e42}) and (\ref{e43}) results in
the wave: $$E=E'+E''.$$ The initial oscillation phases may be
represented as $\sigma_{1}=\eta+\chi$ and $\sigma_{2}=\eta-\chi$.
Then as a result of the wave interference we shall get a rotating
similarity of the standing wave:
\begin{equation}
\label{e44}
E=2A\cos\left(\frac{2\pi}{\mu}\varphi-\chi\right)\exp\left(-i\left(\omega
t-\frac{2\pi}{\lambda}x+\eta\right)\right).
\end{equation}

\begin{figure}[h]
\begin{minipage}[h]{0.32\linewidth}
\center{\includegraphics[width=1\linewidth]{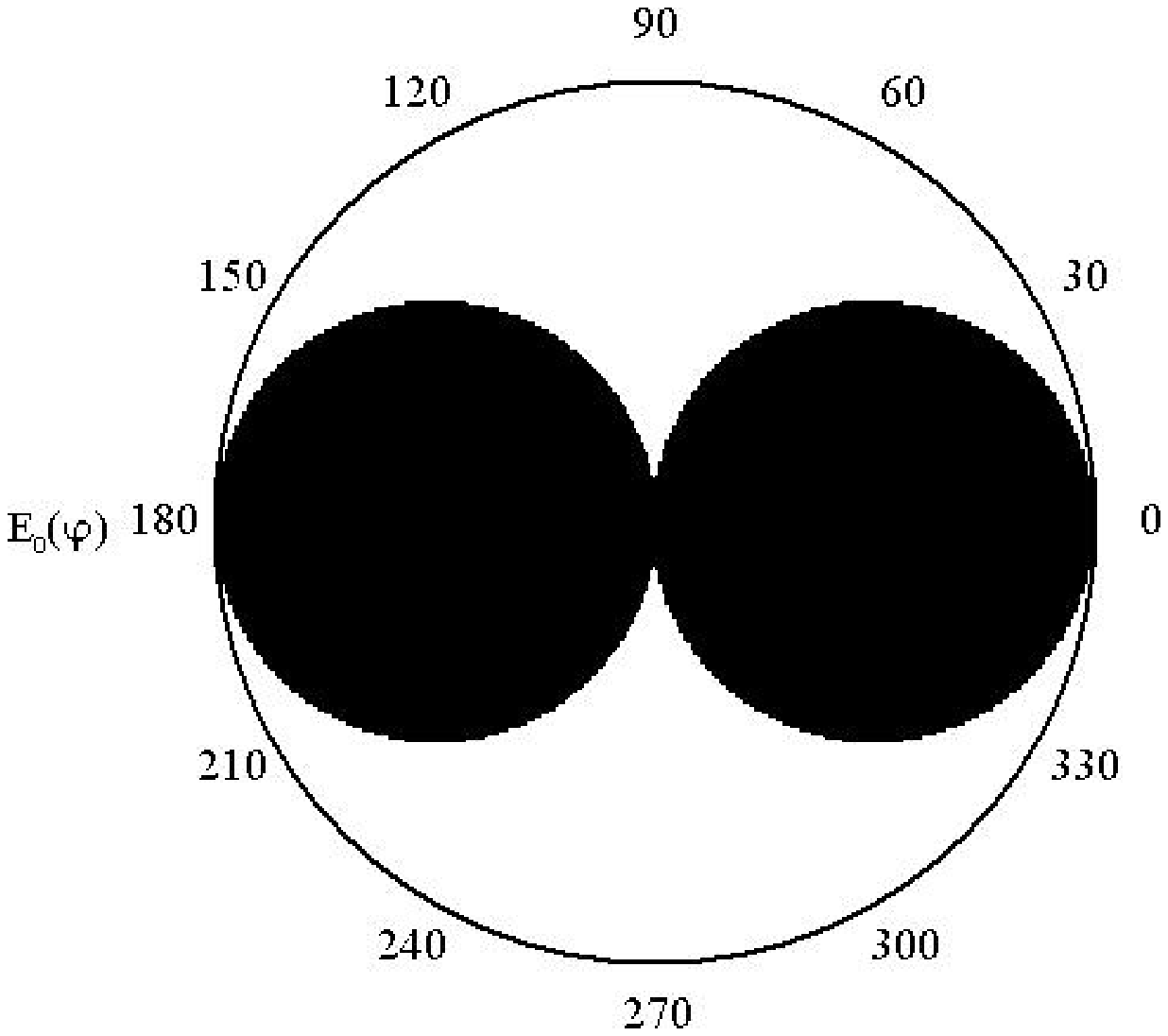}}
\\ a)
\end{minipage}
\hfill
\begin{minipage}[h]{0.32\linewidth}
\center{\includegraphics[width=1\linewidth]{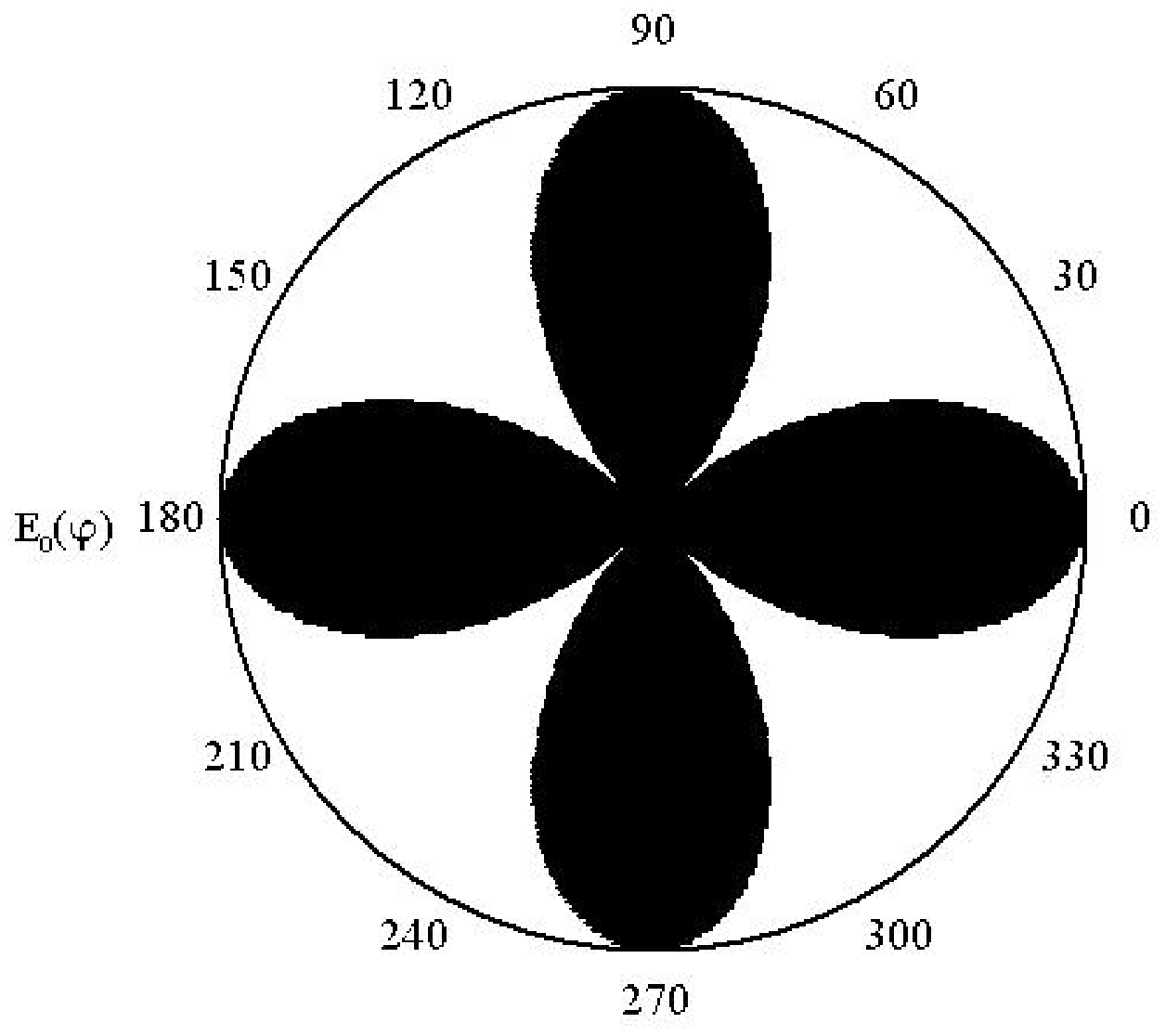}}
\\ b)
\end{minipage}
\hfill
\begin{minipage}[h]{0.32\linewidth}
\center{\includegraphics[width=1\linewidth]{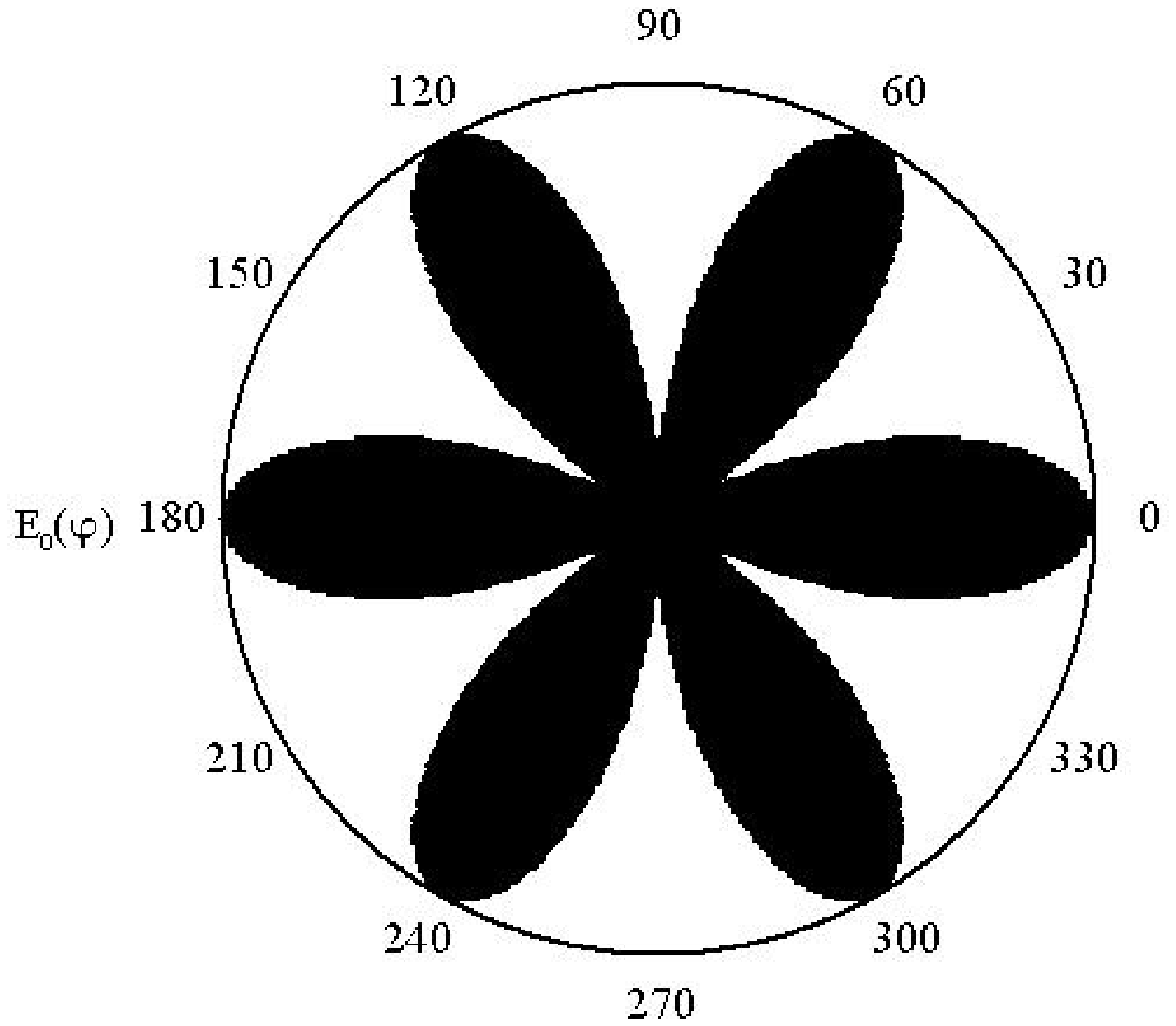}}
\\ c)
\end{minipage}
\caption{At interference of two inversely twisted waves the
amplitude will depend on direction as follows: a) with the
rotation angle $\mu=2\pi$, b) with the rotation angle $\mu=\pi$,
c) with the rotation angle $\mu=2\pi/3$.} \label{fig3}
\end{figure}

From (\ref{e44}) it is seen that if we take:
$$E_{0}(\varphi)=2A\cos\left(\frac{2\pi}{\mu}\varphi-\chi\right)$$
as amplitude of the resulting wave, then the equation (\ref{e44})
reduces to an equation of a standard progressive wave with
amplitude depending on the rotation angle $\varphi$:
$$E=E_{0}(\varphi)\exp\left(-i\left(\omega
t-\frac{2\pi}{\lambda}x+\eta\right)\right).$$

Fig. \ref{fig3} shows the profiles of amplitudes.

\section{Angular Momentum of Twisted Wave}

By substituting (\ref{e39}) to (\ref{e38}) we develop the formula
connecting wavelength, oscillation frequency and rotating
velocity:
\begin{eqnarray*}
\frac{1}{\lambda^{2}}+\frac{1}{\mu^{2}R_{i}^{2}}-\frac{\nu^{2}}{c^{2}}=0.
\end{eqnarray*}
By expressing the wave frequency we obtain dependability of the
frequency on the wavelength and rotation angle in the
gravitational field:
\begin{equation}
\label{e60}
\nu^{2}=\frac{c^{2}}{\lambda^{2}}+\frac{c^{2}}{R_{i}^{2}\mu^{2}}.
\end{equation}

As the work \cite{portnov:bib:Portnov} shows, the Lagrange
function for a free body, in the metric tensor (\ref{e1}), is of
the following form:
\begin{eqnarray*}
L=-mc^{2}\sqrt{1-\frac{V^{2}}{c^{2}}-R_{i}^{2}\frac{\omega^{2}}{c^{2}}},
\end{eqnarray*}
where $m$ - the weight of the body, $V$ - the translational
velocity of the body, $\omega$ - the angular velocity of rotation
of the body.

The impulse and angular momentum of a solid body are traditionally
considered as a vector with components equal to derivatives of the
Lagrange function by the corresponding velocity components. The
impulse of the solid body is equal to:
\begin{eqnarray*}
p^{k}=\frac{mV^{k}}{\sqrt{1-\frac{V^{2}}{c^{2}}-R_{i}^{2}\frac{\omega^{2}}{c^{2}}}},
\end{eqnarray*}
while the angular momentum of the solid body is equal to:
\begin{eqnarray*}
p_{\omega}^{n}=\frac{J_{\omega}\omega^{n}}{\sqrt{1-\frac{V^{2}}{c^{2}}-R_{i}^{2}\frac{\omega^{2}}{c^{2}}}},
\end{eqnarray*}
where $J_{\omega}=m\cdot R_{i}^{2}$ - the inertia momentum of the
body. The energy of the body will be calculated according to the
formula:
\begin{eqnarray*}
E=\frac{mc^{2}}{\sqrt{1-\frac{V^{2}}{c^{2}}-R_{i}^{2}\frac{\omega^{2}}{c^{2}}}}.
\end{eqnarray*}
The total relativistic energy, impulse and angular momentum may be
combined as the seven-dimensional impulse:
\begin{eqnarray*}
P_{7}=\left(\frac{E}{c},p^{k},\frac{p_{\omega}^{n}}{R_{i}^{2}}\right).
\end{eqnarray*}
To develop a formula connecting the energy, impulse and angular
momentum, we shall find the squared 7-impulse of the solid body.
Since the squared 7-velocity is equal to the squared velocity of
light, then the connection between the energy, impulse and angular
momentum will be equal to:
\begin{eqnarray*}
E^{2}=m^{2}c^{4}+p^{2}c^{2}+\frac{p_{\omega}^{2}c^{2}}{R_{i}^{2}}.
\end{eqnarray*}

To describe a photon the connection between the energy, impulse
and angular momentum takes the following form:
\begin{equation}
\label{e61}
E^{2}=p^{2}c^{2}+\frac{p_{\omega}^{2}c^{2}}{R_{i}^{2}}.
\end{equation}
By substituting the energy of photon to $E=h\nu$ and using the
formula (\ref{e60}) we obtain:
\begin{eqnarray*}
p^{2}+\frac{p_{\omega}^{2}}{R_{i}^{2}}=\frac{h^{2}}{\lambda^{2}}+\frac{h^{2}}{R_{i}^{2}\mu^{2}}.
\end{eqnarray*}
The first summand of the right part of the equation is the squared
impulse of photon:
\begin{equation}
\label{e62} p=\frac{h}{\lambda}.
\end{equation}
The second summand is the squared angular momentum of photon:
\begin{equation}
\label{e63} p_{\omega}=\frac{h}{\mu}.
\end{equation}

Let us compare the angular moment of photon derived from the
formula (\ref{e63}) with the orbital angular momentum of impulse
$L_{o}=n\hbar$, derived from the quantum theory. From the formulae
it is seen that quantum number $n$ is the relation of $2\pi$ to
the rotation angle of wave $\mu$:
\begin{eqnarray*}
n=\frac{2\pi}{\mu}.
\end{eqnarray*}
The formula allows to explain the physical meaning of quantum
number $n$. This value is reciprocal of the part of shift of the
wave phase for the period of time, that may be presented as a
relation:
\begin{equation}
\label{e64}
n=\frac{\omega}{\Omega},
\end{equation}
where $\omega$ - the cyclic frequency of electromagnetic wave,
$\Omega$ - the angular velocity of shift of the wave phase.

Let us consider propagation of the electromagnetic wave through a
passive torsional device, for instance, through the pitch-fork
hologram. According to the energy-conservation law the energy of a
non-twisted electromagnetic wave:
\begin{eqnarray*}
E_{1}=\frac{hc}{\lambda_{1}}
\end{eqnarray*}
prior to the propagation through the pitch-fork hologram and the
energy of the twisted electromagnetic wave:
\begin{eqnarray*}
E_{2}=\sqrt{\frac{h^{2}c^{2}}{\lambda_{2}^{2}}+\frac{h^{2}c^{2}}{R_{i}^{2}\mu_{2}^{2}}}
\end{eqnarray*}
following the propagation will be equal: $E_{1}=E_{2}$. Therefore,
the reduced energy-conservation law may be noted as follows:
\begin{equation}
\label{e65}
\frac{1}{\lambda_{1}^{2}}=\frac{1}{\lambda_{2}^{2}}+\frac{1}{R_{i}^{2}\mu_{2}^{2}}.
\end{equation}
As follows from the developed equation (\ref{e65}), there is a new
phenomenon of redshift at twisting light. This phenomenon is that
with twisting of electromagnetic waves by the passive device the
twisted electromagnetic wavelength will be greater compared to the
non-twisted electromagnetic wavelength.

\section{Conclusion}

Let us number the main results of this work obtained within the
framework of generalization of the Maxwell theory of
electromagnetism by seven-dimensional space-time:

1) The equations of twisted light waves within the wave model have
been developed. In particular, the phenomenon of oppositely
twisted light waves has been shown, which, like any phenomenon of
interference, is resulting from the wave nature of light with the
quantitative conformities depending on the wavelength $\lambda$
and the rotation angle $\mu$.

2) From the point of view of wave equations the physical meaning
of the quantum number $n$has been explained, and the formula of
the orbital angular momentum of impulse (\ref{e63}) for the
electromagnetic wave has been developed.

3) The formula of connection of the energy, impulse and orbital
angular momentum for the electromagnetic wave (\ref{e61}) has been
obtained.

4) Based on the energy-conservation law, the new phenomenon of
redshift at twisting of electromagnetic wave has been described.


\begin{thebibliography}{99}

\bibitem{portnov:bib:Allen} L. Allen, M. W. Beijersbergen, R. J. C. Spreeuw, and J. P.
Woerdman, Orbital angular momentum of light and the transformation
of Laguerre-Gaussian laser modes, {\it Phys. Rev. A}, 1992, 45,
8185-8189.

\bibitem{portnov:bib:He} H. He, M. E. J. Friese, N. R. Heckenberg,
and H. Rubinsztein-Dunlop Direct Observation of Transfer of
Angular Momentum to Absorptive Particles from a Laser Beam with a
Phase Singularity, {\it Phys. Rev. Lett.}, 1995, N 75, 826 p.

\bibitem{portnov:bib:Benjamin} Benjamin J. McMorran,
Amit Agrawal, Ian M. Anderson, Andrew A. Herzing, Henri J. Lezec,
Jabez J. McClell and, John Unguris Electron Vortex Beams with High
Quanta of Orbital Angular Momentum, {\it  Science}, 2011, N 6014
Vol. 331, 192--195 pp.

\bibitem{portnov:bib:Arita} Yoshihiko Arita,
Michael Mazilu, Kishan Dholakia Laser-induced rotation and cooling
of a trapped microgyroscope in vacuum, {\it Nature
Communications}, 2013, N 4, 3--11 pp.

\bibitem{portnov:bib:Bahrdt} J.
Bahrdt, K. Holldack, P. Kuske, R. Muller, M. Scheer, and P. Schmid
First Observation of Photons Carrying Orbital Angular Momentum in
Undulator Radiation, {\it Phys. Rev. Lett.}, 2013, N 111, 034801.

\bibitem{portnov:bib:Torres} Juan P. Torres, Lluis Torner {\it
Twisted Photons: Applications of Light with Orbital Angular
Momentum} (Wiley-VCH: 2011), 288 p.

\bibitem{portnov:bib:Tamburini} F. Tamburini, B. Thide, G. Molina-Terriza, G.
Anzolin, Twisting light around rotating black holes, {\it Nature
Physics}, 2011, No. 7 (3), pp. 195-197. arXiv:1104.3099.

\bibitem{portnov:bib:BabLipFrol} Baburova O.V., Lipkin K.N., Frolov B.N.
Theory of gravity with the Dirac scalar field and the problem of
cosmological constant, {\it Russian Physics Journal}, 2012, V.55,
N.7, pp. 855-857.

\bibitem{portnov:bib:PavPan} V. N. Pavelkin, V. F. Panov A nonstationary cosmological model
with rotation in the Einstein-Cartan theory, {\it Russian Physics
Journal}, August 1993, Volume 36, Issue 8, pp. 784-788

\bibitem{portnov:bib:KufPan} E. V. Kufshinova, V. F. Panov Quantum Birth of the Rotating
Universe, {\it Russian Physics Journal}, October 2003, Volume 46,
Issue 10, pp. 999-1009.

\bibitem{portnov:bib:Krech} Krechet V. G., Sadovnikov D. V. Cosmology in an affine-metric
theory of gravity with a scalar field, {\it Gravitation and
Cosmology}, 1997, V.3, pp. 133-140.

\bibitem{portnov:bib:BabFrol} Babourova O. V., Frolov B. N. Colour-spin, dilaton-spin and
hypermomentum perfect fluids as the sources of non-Riemannian
cosmologies, {\it Nucl. Phys. B}, Proc. Suppl. (Proc. 19th Texas
Symp. on Relativistic Astrophysics and Cosmology, Paris, 1998),
2000, V.80, pp. 04/01 1 9.

\bibitem{portnov:bib:Teyssandler} Teyssandier P., Tucker R. W. and Wang C. On an interpretation of
non-Riemannian gravitation, {\it Acta Phys. Polonica B}, 1998, V.
29, pp. 987-994.

\bibitem{portnov:bib:Puetzfeld} Puetzfeld D. A cosmological model in Weyl-Cartan spacetime: I.
Field equations and solutions, {\it Class. Quantum Grav.}, 2002,
V. 19, pp. 3263-3280 (gr-qc/0111014).

\bibitem{portnov:bib:Miritzis} Miritzis J. Isotropic cosmologies in Weyl geometry, {\it Class.
Quantum Grav.}, 2004, V. 21, pp. 3044-3056 (gr-qc/0402039).

\bibitem{portnov:bib:Portnov2006} Portnov Yu. A. Dilatation field quanta, {\it Gravitation and Cosmology},
2006, V.12, pp. 209-211.

\bibitem{portnov:bib:Portnov2015} Portnov Yu.A. Descending entropy in expanding the universe, {\it
International Journal of Geometric Methods in Modern Physics},
2015, Vol. 12, 1550024 (5 pages)

\bibitem{portnov:bib:Portnov} Yu. A. Portnov, {\it The Field Equations in
Seven-Dimensional Space-Time} (Moscow State University of Printing
Arts, Moscow, 2013), p. 154.

\bibitem{portnov:bib:PortnovGC} Portnov~Y.\,A. Gravitational Interaction in
Seven-Dimensional Space-Time, {\it Gravitation and Cosmology},
2011, N 17, 152--160 pp.

\bibitem{portnov:bib:PortnovLD} Portnov~Y.\,A. On variation in spin rate of bodies in a variable
gravity field, {\it Annales de la Fondation Louis de Broglie},
2014, Vol. 39, pp. 89-99.

\bibitem{portnov:bib:PortnovRAP} Portnov~Y.\,A. Gravity Probe B Experiment
in 7D Space-and-time Continuum, {\it Review of Applied Physics},
2013, N 4, 96--98 pp., arXiv:1204.5175v1.

\bibitem{portnov:bib:PortnovGC2014} Portnov~Y.\,A. Obtaining galaxy rotation curves without dark matter, {\it Gravitation and Cosmology},
2014, Vol. 20, No. 4, pp. 279-281.

\bibitem{portnov:bib:PortnovST2014} Portnov~Y.\,A. Dynamic and static cosmological constant in seven-demensional gravitational equations, {\it Space, time and fundamental interactions}, 2014, No. 3, pp. 32-37.

\bibitem{portnov:bib:Hermann} Robert Hermann {\it Ricci and
Levi-Civita's Tensor Analysis Paper} (Brookline: Math Sci Press,
1975), 273~p.

\bibitem{portnov:bib:Dimitrienko} Dimitrienko Yu.I. {\it The tensor calculus} (High school, Moscow,
2001), 575 p.


\end{thebibliography}
\end{document}